\documentclass[aps,prb,superscriptaddress,reprint,amsmath,amssymb]{revtex4-2}
\usepackage{color}
\usepackage{graphicx}% Include figure files
\usepackage{dcolumn}% Align table columns on decimal point
\usepackage{bm}% bold math
\usepackage[mathlines]{lineno}% Enable numbering of text and display math
\usepackage{tabularx}
\usepackage[colorlinks,linkcolor=blue,anchorcolor=blue,citecolor=blue,urlcolor=blue,hyperindex,CJKbookmarks]{hyperref}
\usepackage{footnote}
\usepackage{float} 
\usepackage{amsmath}
\usepackage{txfonts}
\usepackage{mathptmx}
\usepackage{ragged2e}
\usepackage{booktabs,makecell, multirow, tabularx}
\usepackage{multirow}
\usepackage{braket}
\usepackage{gensymb}
\usepackage{array}
\usepackage{lmodern} 
\usepackage{fix-cm} 
\usepackage{textcomp} 
\begin{document}

%\preprint{}

\title{A DFT+DMFT study of the electronic structure of Samarium}


\author{Shengsong Xu}\affiliation{School of Physics and Beijing Key Laboratory of Opto-electronic Functional Materials $\&$ Micro-nano Devices, Renmin University of China, Beijing 100872, China}\affiliation{Key Laboratory of Quantum State Construction and Manipulation (Ministry of Education), Renmin University of China, Beijing 100872, China}
\author{Zhenfeng Ouyang}\affiliation{School of Physics and Beijing Key Laboratory of Opto-electronic Functional Materials $\&$ Micro-nano Devices, Renmin University of China, Beijing 100872, China}\affiliation{Key Laboratory of Quantum State Construction and Manipulation (Ministry of Education), Renmin University of China, Beijing 100872, China}
\author{Li Huang}\email{huangli@caep.cn}\affiliation{National Key Laboratory of Surface Physics and Chemistry, Mianyang 621908, China}
\author{Zhong-Yi Lu}\email{zlu@ruc.edu.cn}\affiliation{School of Physics and Beijing Key Laboratory of Opto-electronic Functional Materials $\&$ Micro-nano Devices, Renmin University of China, Beijing 100872, China}\affiliation{Key Laboratory of Quantum State Construction and Manipulation (Ministry of Education), Renmin University of China, Beijing 100872, China}\affiliation{Hefei National Laboratory, Hefei 230088, China}
%\thanks{}
%\affiliation{}

\date{\today}

\begin{abstract}
The electronic structure of Samarium (Sm) was calculated using the density functional theory combined with the single-site dynamical mean-field theory. In this work, we investigated
the electronic properties of $\alpha$, $\beta$ and $\gamma$ phases at ambient pressure, including the band structures, density of states, self-energy functions and valence state histograms. 
Our results agree with the experimental data.The calculation shows that the $4f$ electrons in all these phases are well localized, the Kondo peaks are
suppressed and the hybridization between the $4f$ electrons and conduction electrons are quite weak. Our results also show the strong correlation effect is significant in Sm metal.

\end{abstract}

%\keywords{}

\maketitle

\section{Introduction}

%Outline: Sm struct  /  4f  electrons  / afm mag / 

Samarium (Sm) is a fascinating lanthanide element with atomic number 62, exhibiting a rich temperature-pressure phase diagram which includes several crystallographic phases and antiferromagnetic (AFM) phases 
at low temperatures. As temperature increases from 0 K at ambient pressure, Sm undergoes three structural phases in sequence \cite{Daane:a01231,MARDON1970477,COLES1981153}: a 
peculiar rhombohedral 9R structure ($\alpha$ phase), a hexagonal close-packed (hcp) structure ($\beta$ phase) and a body-centered cubic (bcc) structure ($\gamma$ phase). With increasing pressure, 
a series of phase transitions of Sm type (hR9) $\rightarrow$ dhcp (hP4) $\rightarrow$ fcc (cF4) $\rightarrow$ distorted fcc will occur \cite{PhysRevB.101.174109}. This structural transition sequence 
is observed to be shared by many lanthanide elements neighboring Sm in the periodic table \cite{PhysRevB.101.174109}.

Samarium also displays complex magnetic properties due to the partially filled 4$f$ electron shell. At ambient pressure, the $\alpha$ phase of Sm exhibits two AFM phases,
the low temperature phase exists below 14 K while the high temperature phase exists between 14 K and 106 K. Due to the low symmetry of the $\alpha$ phase, the spin structures of AFM phases
are quite complex\cite{PhysRevLett.29.1468}. Along with the structural transitions by increasing pressure, the high temperature AFM phase disappears at 13 GPa, while the low temperature AFM phase 
persists in the sequence of high pressure structural phases \cite{PhysRevB.99.085137}.

Sm owns many interesting compounds with fascinating properties. $\mathrm{SmCo_5}$ and 
$\mathrm{Sm_2Co_{17}}$, are brilliant permanent magnets \cite{10.1063.1.1709459}. This makes Sm an important element in the field of magnetic materials. $\mathrm{SmB_6}$ is a 
paramagnetic bulk insulator that shows non-trivial topological properties \cite{Min2017,Hatnean2013,PhysRevLett.112.136401}. $\mathrm{SmS}$ produces the hybridization-induced pseudogap feature, together with temperature-dependent and pressure-dependent
behaviors of electronic structure and resistivity \cite{PhysRevLett.114.166404,PhysRevB.105.195135}. Carbon-boron clathrate of Sm shows strong Kondo lattice behavior at low temperature \cite{PhysRevB.106.245131}. 
The singular properties of these compounds mainly originate from the strongly correlated $4f$ electrons of Sm, whose atomic configuration is $[Xe]4f^65d^06s^2$.

However, the electronic properties of Sm metal are not fully understood yet. The main challenge also comes from the strongly correlation properties $4f$ electrons. 
In this work, we employed a first-principles many-body approach, the density functional theory combined with the dynamical mean-field theory (DFT + DMFT) to study the 
electronic structures of the three phases of Sm metal at ambient pressure. Many lanthanides \cite{PhysRevLett.87.276403,PhysRevLett.87.276404,PhysRevB.99.045122,PhysRevB.80.235105}, like Ce, Pr, Nd, and actinides \cite{Shim2007,PhysRevB.94.115148,PhysRevB.101.125123}, like U, Np, Pu, Am, Cm, had been well studied by this method, 
which shows the effectiveness of DFT + DMFT in the study of correlated f electrons. We calculated the fundamental observables of electronic structure including the spectral functions, the total and $4f$ partial
density of states. Also we investigated the strong correlation properties like the self-energy functions and the valence state histogram.% Our calculation shows that

The general organization of this article is as follows: In Sec. II, we briefly introduce the framework of the DFT + DMFT method. In Sec. III, we present our calculation results in several subsections 
and give discussions. At last, we give a brief conclusion in Sec. IV.

\section{Method}

The $4f$ electrons in lanthanide elements are strongly correlated that they can 
not be properly treated by the density functional theory (DFT) based electronic structure calculation \cite{RevModPhys.78.865}, which is based on the independent-electron picture. To take the strong correlation effect into account, the single-site dynamical mean-field theory (DMFT)
is combined with DFT to form the DFT + DMFT method \cite{VAnisimov1997,PhysRevB.57.6884,RevModPhys.78.865}. DMFT is a nonperturbative many-body approach that maps lattice models onto the Anderson impurity model, thereby including local quantum fluctuations 
and producing a momentum‑independent self‑energy \cite{RevModPhys.68.13}. This mapping becomes exact in the limit of infinite dimensions, but also works well in three-dimensions
\cite{PhysRevLett.62.324,MllerHartmann1989,RevModPhys.68.13}.  In this calculation, the DFT part was performed using the WIEN2K program \cite{wien2k} and the DMFT part was performed by the EDMFTF package \cite{PhysRevB.81.195107}. 

\begin{figure}[t]
\centering
\includegraphics[width=0.5\textwidth]{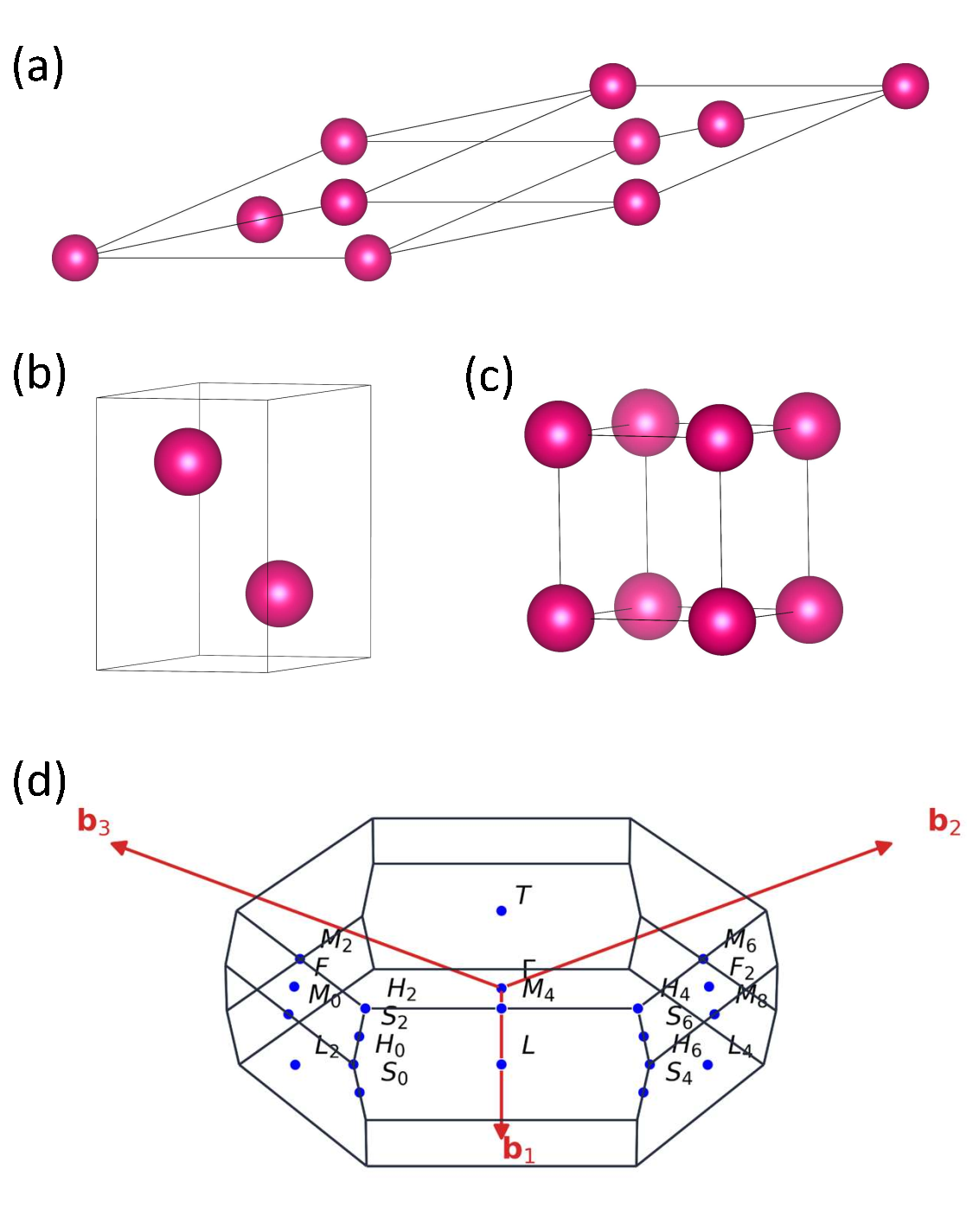}
\caption{\label{} Crystal structures of the three phases of Sm metal. (a) rhombohedral structure ($\alpha$ phase). (b) hexagonal close-packed structure ($\beta$ phase). (c) body-centered cubic structure ($\gamma$ phase). (d) the Brillouin zone of the rhombohedral structure. The high symmetry points are labeled. \cite{HINUMA2017140,togo2024}}
\end{figure}

In the DFT part, we used the crystal structures measured in experiments for the $\alpha$ and $\beta$  phases \cite{Daane:a01231,Kumar:a07033}. While for the $\gamma$ phase, the structure was obtained by
structural relaxation using VASP \cite{PhysRevB.54.11169} due to the lack of experimental data, we derived the lattice constant $a = 3.253 \AA$. The WIEN2K program uses the full-potential linear augmented plane-wave formalism (FP-LAPW),
and in the calculation of the exchange-correlation potential, it uses the Perdew-Burke-Ernzerhof functional \cite{PhysRevLett.77.3865}. The k-points meshes of the bcc and rhombohedral structures  
were $17 \times 17 \times 17$ and the hcp structure was $20 \times 20 \times 11$. The muffin-tin radius $R_{MT}$ of Sm was set to 2.50 Bohr, and the plane-wave cutoff was determined by 
$R_{MT}K_{max}$ = 7.0. The spin-orbit coupling effect was included in all calculations.

The DMFT part was performed by the EDMFTF package with the hybridization expansion continuous-time quantum Monte Carlo (CT-QMC) method as the impurity solver \cite{PhysRevLett.97.076405,PhysRevB.75.155113,RevModPhys.83.349}. The correlated orbitals were 
chosen to be the 4$f$ orbitals of Sm, and the on-site Coulomb interaction $U$ and Hund's exchange $J_H$ were set to 6.1 eV and 0.835 eV respectively \cite{PhysRevB.98.075129}. For the $\beta$  and $\gamma$ phases, the temperature was set to 1100 K 
and 1300K, which lie in the middle of their temperature ranges on the phase diagram at ambient pressure \cite{COLES1981153}. For the $\alpha$ phase, we picked up several temperatures from 120 K
to 600 K. The AFM exists below 106 K in the $\alpha$ phase, so it is not considered in this work. We performed charge fully self-consistent DFT + DMFT calculations. About 40-60 iterations are enough to achieve good convergences in charge densities, 
self-energy functions, and chemical potentials for the three ambient phases of Sm. The number of Monte Carlo steps in the CT-QMC impurity solver was set to $1 \times 10^8$ per parallel processor. 
As CT-QMC is done on imaginary axis, the maximum entropy method is used to accomplish the analytical continuation to derive the Green's function on real axis, which gives the density of states \cite{JARRELL1996133}. Another point is 
the impurity occupancies, as the $f$-electron has a large many-body Hilbert space, it is necessary to truncate the atomic eigenstates to a smaller subspace by setting the occupancy to a certain range. 
The nominal occupancy of 4f shell in Sm is $N = 6$, while in our calculations we found that the atomic eigenstates actually well concentrated in the $N=5$ subspace. So we kept $N \in \left[3, 7\right]$ 
in the $\beta$  and $\gamma$ phases, and in the $\alpha$ phase we set $N \in \left[4, 6\right]$ for a balance between accuracy and computational cost. 

\begin{figure*}[t]
\centering
\includegraphics[width=\textwidth]{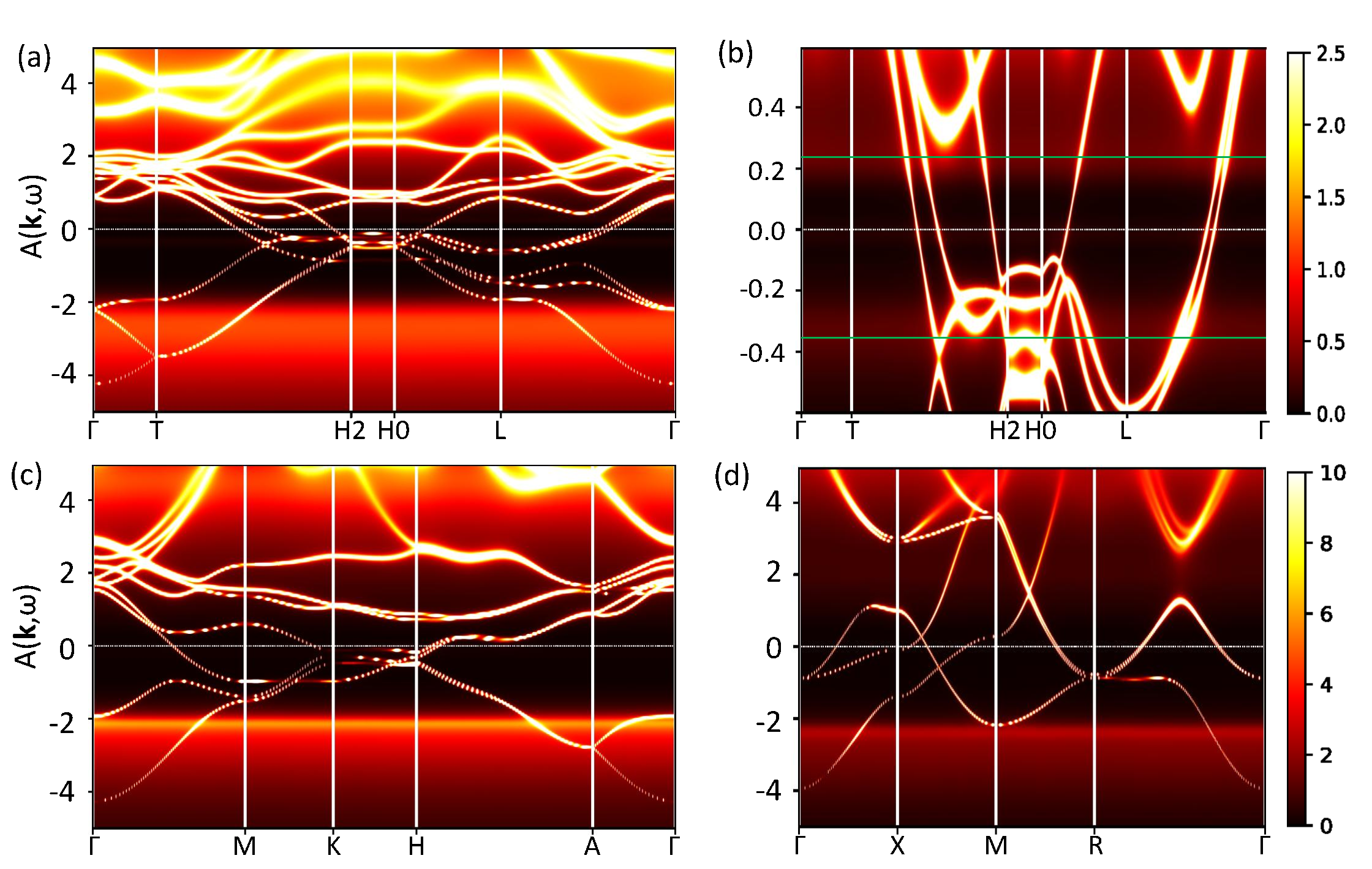}
\caption{\label{}  Momentum-resolved spectral functions $A(\mathbf{k},\omega)$ of the Sm metal. The (a) (c) (d) shares the same colorbar. (a) the $\alpha$ phase at 120 K. (b) the $\alpha$ phase at the interval of $[-0.6,0.6]$ around Fermi level, the colorbar is specifically adjusted to make the flat bands stand out. The centers of the flat bands are labeled by green solid lines.
There are two flat bands above and below the Fermi level. (c) the $\beta$ phase at 1100 K. (d) the $\gamma$ phase at 1300 K. }
\end{figure*}

\section{Results and discussion}
\subsection{Band structures}
In the DFT + DMFT method, it is the momentum-resolved spectral function $A(\mathbf{k},\omega)$ that plays the role of the band structure $E(k)$ in DFT. The spectral function reflects the quasiparticle band 
structure, thus is a fundamental theoretical tool in the study of materials. The calculated spectral functions along high symmetry lines are illustrated in Fig. 2, in an energy window of $[-5, 5]$ eV.

Examining the presented spectral pictures in Fig. 2, we notice several remarkable features. First, in the $\alpha$ phase, the Kondo resonance peaks, which are the characteristic feature of $4f$ electrons \cite{PhysRevB.99.045122,PhysRevLett.112.136401},
are suppressed and only reserve very weak flat band. This is clearly zoomed in Fig. 2(b) with the corresponding density of states in Fig. 3(e). This is in sharp  contrast to the Kondo resonance compounds like $\mathrm{SmB_6}$ and  SmS \cite{PhysRevLett.114.166404,PhysRevLett.112.136401}. 
Second, the lower and upper Hubbard bands of $4f$ electrons in the $\alpha$ phase respectively centered at -2 eV and 5 eV are quite blurry, indicating their incoherence. In contrast, 
the Hubbard bands of the high temperature phases are more coherent. Third, besides $4f$ electrons, there are still other noticeable features. All the phases have conduction bands crossing the Fermi level, showing their metallic nature.
The bands of the $\alpha$ phase between the high symmetry points H0 and H2 at the interval of -2 eV to 0 eV are flat, this feature comes from the symmetry of the rhombohedral structure.

As the ground state of Sm metal has AFM order, it is considered due to the Ruderman-Kittel-Kasuya-Yosida (RKKY) interaction \cite{PhysRevLett.29.1468}. The RKKY interaction competes with the Kondo
resonance effect \cite{DONIACH1977231}, even above the $\mathrm{N\acute{e}el}$ temperature, it still disturbs the forming of the Kondo resonance peak. Of course, the 120 K is also not a low temperature, which is 
another factor of the weak Kondo resonance peak. The weak Kondo resonance in these phases derives a small $c-f$ hybridization as there is insufficient spectral weight of $4f$ electrons
near the Fermi level to hybridize with conduction electrons, indicating the localized nature of the $4f$ electrons.    

\begin{figure*}[t]
\centering
\includegraphics[width=\textwidth]{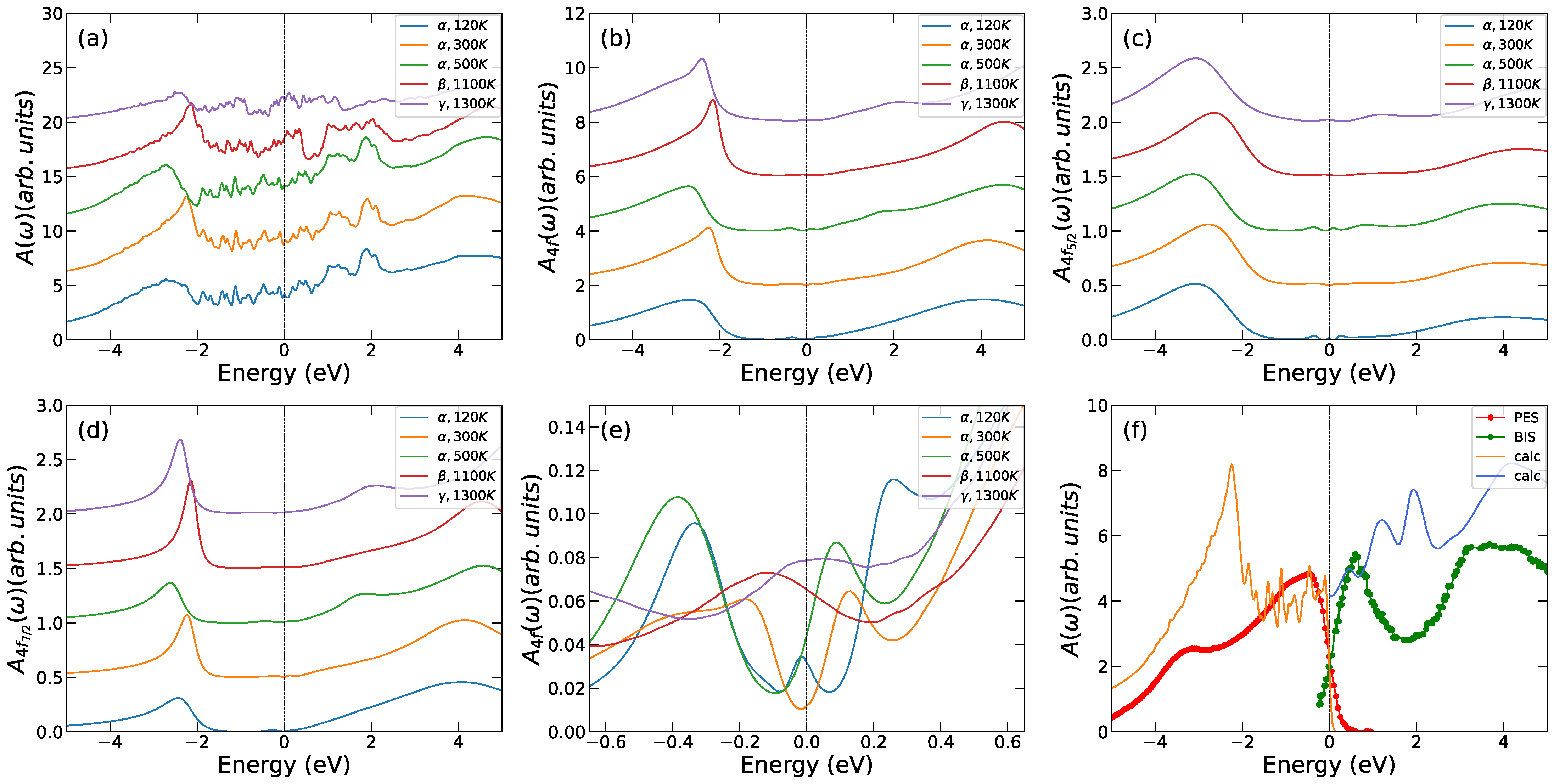}
\caption{\label{}  Density of states of Sm by DFT+DMFT calculations. (a) Total density of states $A(\omega)$. (b) $4f$ partial density of states $A_{4f}(\omega)$ 
 (c) Orbital-resolved $4f$ partial density of states $A_{4f_{5/2}}(\omega)$. (d) Orbital-resolved $4f$ partial density of states $A_{4f_{7/2}}(\omega)$. (e) $4f$ partial density of states 
$A_{4f}(\omega)$ near the Fermi level. (f) Total density of states $A(\omega)$ at 300 K, compared with the experimental UV-photoemission spectrum (PES) data \cite{Brodén1972} and Bremsstrahlung isochromat spectroscopy (BIS) \cite{LANG1979945}.}
\end{figure*}

\subsection{Density of states}

The density of states also provides detailed information about the material. We present the many kinds of density of states in Fig. 3, including the total density of states $A(\omega)$ and the
partial density of states of $4f$ electrons $A_{4f}(\omega)$. The $4f$ orbitals, considering the spin-orbit coupling effect, are split into $A_{4f_{5/2}}(\omega)$ and $A_{4f_{7/2}}(\omega)$
states. The corresponding density of states of both orbitals are presented. We use the energy interval $[-5,5]$ eV to contain the lower and upper Hubbard bands.

Comparing the density of states of different phases at different temperatures, we see many interesting features. At first, the density of states of $4f$ electrons is concentrated in 
the lower and upper Hubbard bands, and this contribution can be seen in the the total density of states. Second, it is obvious that the density of states near the Fermi level mainly comes from the
conduction electrons, and the Kondo peaks of $4f$ electrons are suppressed. But the Kondo peaks do exist in the $\alpha$ phase as we show in Fig. 3 (e),even though they are really weak.
In comparison, the densities of states of the $\beta$ and $\gamma$ phases don't exhibit this structure. The Kondo peaks in the $\alpha$ correspond to the weak flat bands in Fig. 2 (b). 
These features mentioned above again give the picture that the $4f$ electrons mainly stay at the lower Hubbard band at -2.5 eV, so they are localized and have little hybridization with the
conduction electrons. Third, the temperature dependent variations of density of states are noticeable. The height and width of the peak corresponding to the lower Hubbard band vary with the temperature,
but its position fixed. The shape of the upper Hubbard band is stable but its position slightly shift to higher energy as temperature increases. The Kondo peaks also change at different temperature.

The calculated density of states can be compared with experiment results. To be comparable with the experimental data, first, the part of $A(\omega)$ under Fermi level should be multiplied by a Fermi-Dirac distribution 
function, and the upper part need to be broadened using a Gaussian function \cite{PhysRevB.30.6921}. An ultraviolet photoemission experiment gave energy distribution curves (EDC) which reflects the density of states under 
the Fermi level, here we choose the EDC measured at incident photon energy of 7.4 eV \cite{Brodén1972}. A Bremsstrahlung isochromat spectroscopy (BIS) experiment gave spectrum reflecting the density
of states above the Fermi level \cite{LANG1979945}. We compare the experiment data with the calculated density of states of the $\alpha$ phase at room temperature in Fig. 3(f). Under the Fermi level, 
the wide peak in experiment data can be regarded as a fusion of the two calculated peaks at -0.5 eV and -1.1 eV. And the small peak at -3.2eV may correspond to the lower Hubbard band. 
Above the Fermi level, the calculated peak at 0.5 eV is captured by the BIS data, together with the upper Hubbard band at 4 eV. However, the calculated peaks at 1 eV and 2 eV do not exist
in the BIS data. Our calculations roughly capture some signs in experiments. The discrepancies between the theoretical and experimental spectrum may come from the approximations used in the DFT + 
DMFT method, and technically the uncertainty in the Coulomb interaction parameters and the usage of oversimplified double-counting scheme \cite{Anisimov_1997}.

\subsection{Self-energy functions}

The self-energy functions contain the important information on electronic correlation. The Matsubara self-energy functions of the $4f_{5/2}$ and
$4f_{7/2}$ components are presented in Fig  4 and show interesting features. The concave curves represent the metallic feature, which is the
common feature of most cases, including the low temperature region of the $\alpha$ phase, and the $\beta$ phase of the $4f_{5/2}$ orbital, 
and all of the $4f_{7/2}$ orbital. The high temperature region of the $\alpha$ phase and the $\gamma$ phase of the $4f_{5/2}$ orbital show insulator 
feature. The intercept in the y-axis represents the scattering rate $\gamma_s$, it is obvious that the scattering rate of the $4f_{5/2}$ orbital is larger than
the $4f_{7/2}$ orbital. These observations show the differences between the two orbitals.

The imaginary part of low-frequency Matsubara self-energy functions can be fitted in the following form:
\begin{equation}
    -\mathrm{Im} \Sigma(i\omega_n) = A ( i\omega_n)^\alpha + \gamma_s
\end{equation}
The exponent fitting parameter $\alpha$ serves as an indicator for the Fermi liquid behavior. $\alpha = 1$ is the character of Landau Fermi liquid. 
In Fig. 4, the curves near the zero frequency are far from the linear behavior, which indicates non-Fermi liquid behavior.     

\begin{figure}[thb]
\centering
\includegraphics[width=0.5\textwidth]{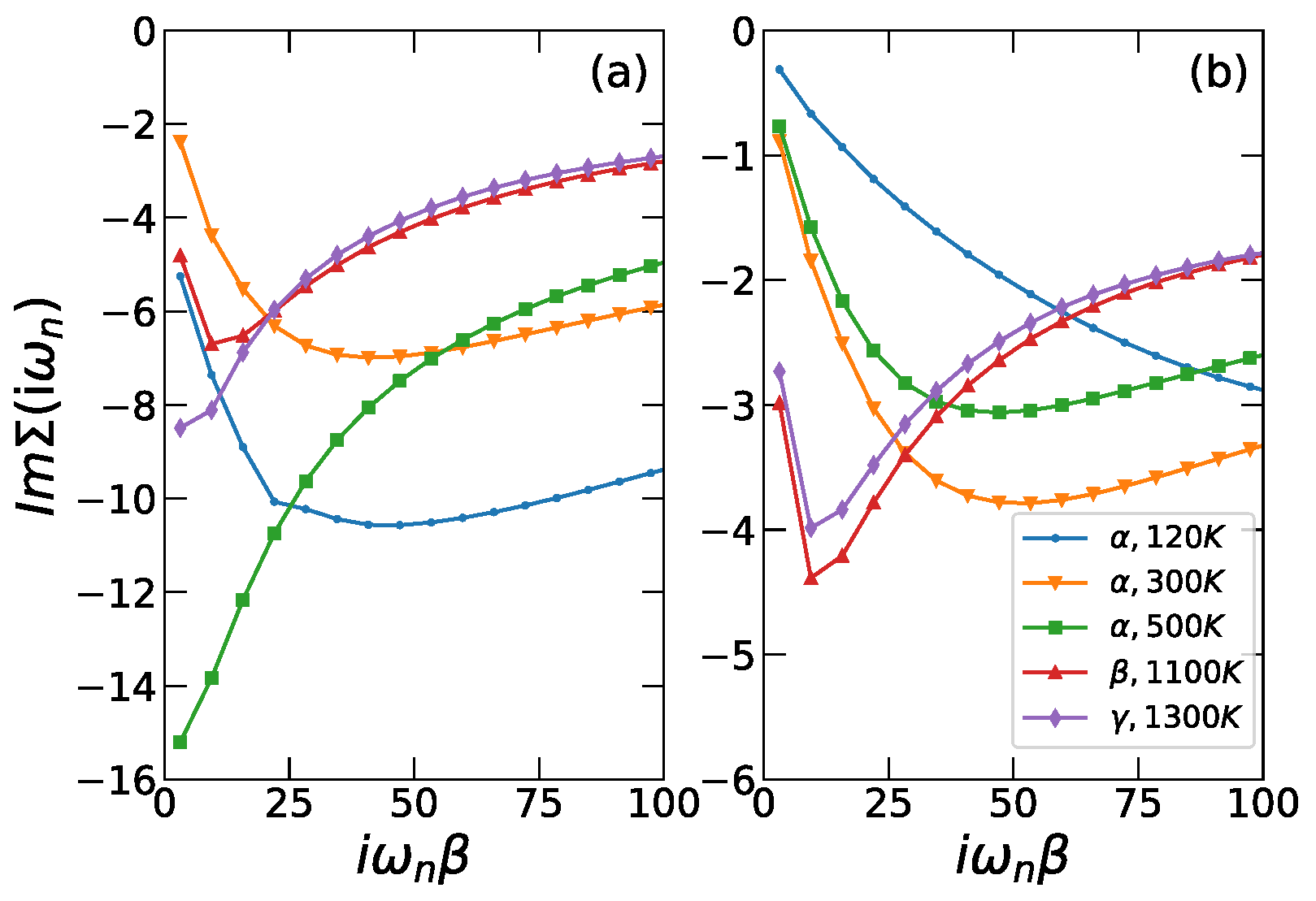}
\caption{\label{}Imaginary parts of the Matsubara self-energy functions of Sm at low-frequency by DFT + DMFT calculations. (a) $4f_{5/2}$ components. (b) $4f_{7/2}$ components.  }
\end{figure}

Another important quantity can be extracted from the self-energy functions is the quasiparticle weight $Z$, and the associated effective mass $m^*$ 
\cite{RevModPhys.68.13}.
\begin{equation}
    Z^{-1} = \frac{m^*}{m_e}  \approx 1 - \frac{\partial \mathrm{Im} \Sigma(i\omega_0)}{\partial i\omega_0} 
\end{equation}
Here $\omega_0 = \pi / \beta$ and $m_e$ denotes the mass of an electron. The calculated quasiparticle weights and effective masses are listed 
in Table I. The small quasiparticle weights show the very strength of the strong correlation effect and band renormalization of $4f$ electrons in Sm metal, they are smaller than Sm compound \cite{PhysRevB.105.195135}. The orbital difference
is also notable, the $4f_{5/2}$ orbital has a larger effective mass and a smaller quasiparticle weight than the $4f_{7/2}$ orbital. The temperature 
dependence of the effective mass of the $4f_{5/2}$ orbital in the $\alpha$ phase is an interesting feature. We think it is a result of the combined
effect of the thermal fluctuations and the RKKY interaction.  

\begin{table}[b]
\caption{\label{}Calculated quasiparticle weights Z and electron effective masses $m^*$ of Sm, the unit of effective masses $m^*$ is electron 
mass $m_e$.}
\begin{ruledtabular}
\begin{tabular}{rcccc}
        Cases & $m^*_{5/2}/m_e$ & $Z_{5/2}$ & $m^*_{7/2}/m_e$ & $Z_{7/2}$  \\ \hline
        120K & 162.72 & $6.14*10^{-3}$& 10.63 & $9.41*10^{-2}$  \\
        300K & 30.21 & $3.31*10^{-2}$ & 11.91 & $8.39*10^{-2}$  \\ 
        500K & 113.36 & $8.82*10^{-3}$ & 6.6587 & $1.50*10^{-1}$  \\ 
        1100K & 17.16 & $5.82*10^{-2}$ & 11.03 & $9.06*10^{-2}$ \\ 
        1300K & 25.13 & $3.98*10^{-2}$ & 8.765 & $1.14*10^{-1}$ \\ 
\end{tabular}
\end{ruledtabular}
\end{table}

\subsection{Valence state histograms}
The CTQMC impurity solver records the probability distribution of the atomic eigenstates, which contains valuable information about the valence
configuration. Summing the probabilities that lie in a subspace of the Hilbert space with certain quantum numbers, the occupancy number $N$ and the total 
angular momentum $J$, yields the valence state histogram in Fig. 5. 
\begin{figure}[t]
\centering
\includegraphics[width=0.5\textwidth]{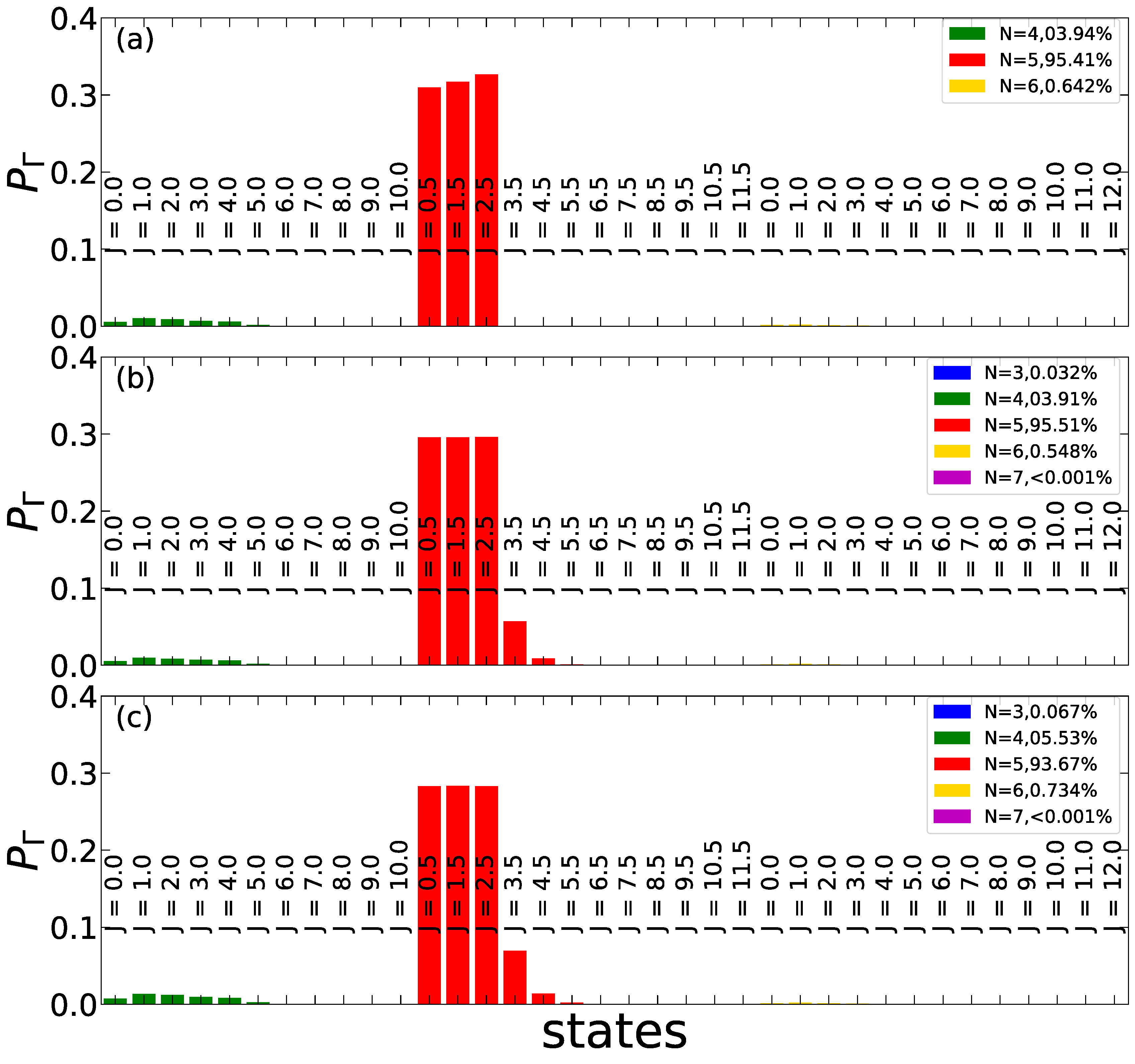}
\caption{\label{} Valence state histograms of Sm by DFT + DMFT calculations. The atomic eigenstates are labeled by good quantum numbers N 
(total occupancy number) and J (total angular momentum).(a) the $\alpha$ phase at 120 K. (b) the $\beta$ phase at 1100 K. (c) the $\gamma$ phase
 at 1300 K. The probability distribution of $N=3,N=7$ states are so small that they are omitted in the figure.}
\end{figure}
From the histograms, it is obvious that the $N=5$ occupancy number dominates the probability distribution with a share around $95\%$. This $4f$ valence state widely exists in Sm compounds \cite{PhysRevLett.114.166404,PhysRevLett.112.136401,PhysRevB.106.245131}.
The distribution of $N = 4, N=6$ subspaces is at an order of $1\%$. The $N=3, N=7$ subspaces are counted in the calculation of $\beta$ and $\gamma$ phases, and
the result shows they are negligible. This picture is in contrast to the corresponding $5f$ element Pu with atomic configuration $[Rn]5f^66d^07s^2$ \cite{PhysRevB.101.125123}, only $55\%$ of 
the valence state of which concentrates on $N=5$. As the valence state fluctuation of Pu is stronger, it has weaker correlation and it is closer to a Fermi liquid. For Sm, the concentration of the probability distribution in the $N=5$ subspace is consistent with the localized picture
of $4f$ electrons. Within the $N=5$ subspace, the total angular momentum $J$ mainly distributes in three values $J=0.5,1.5,2.5$. The angular momentum distributions
of the three phases are highly similar, but there are still some differences. The probability distribution among the three $J$ values is less equivalent in the
$\alpha$ phase, while in the $\beta$ and $\gamma$ phases, the $J=3.5,4.5$ has a small contribution. 

\section{Conclusion}

In the present work, we performed DFT+DMFT calculations to study the electronic structures of the three phases of Samarium metal at ambient pressure. 
We calculated the momentum-resolved spectral functions, the total and $4f$ partial density of states, the self-energy functions and the valence state histograms.
Our results indicate that the $4f$ electrons are well localized with little hybridization with the conduction electrons. The Kondo resonance peak of $4f$ electrons is
suppressed, and the density of states near the Fermi level mainly comes from the conduction electrons. The calculated density of states capture the features in 
the experimental data. The system exhibits non-Fermi liquid behavior, and the large effective masses of $4f$ electrons show that the strong correlation effect is significant. 
Our work provides a comprehensive picture of the electronic structure of Sm metal, and will be helpful for the understanding of $4f$ electrons in lanthanide elements and their compounds.

\begin{acknowledgments}
This work was supported by National Key R$\&$D Program of China (Grants No. 2024YFA1408601, Grants No. 2024YFA1408602), the
National Natural Science Foundation of China (Grant No. 12434009, No. 12274380, No. T2525034), and the Innovation Program for Quantum Science and Technology (Grant No.2021ZD0302402). 
Computational resources were provided by the Physical Laboratory of High Performance Computing at Renmin University of China.
\end{acknowledgments}

\bibliography{Sm}

\begin{thebibliography}{43}%
\makeatletter
\providecommand \@ifxundefined [1]{%
 \@ifx{#1\undefined}
}%
\providecommand \@ifnum [1]{%
 \ifnum #1\expandafter \@firstoftwo
 \else \expandafter \@secondoftwo
 \fi
}%
\providecommand \@ifx [1]{%
 \ifx #1\expandafter \@firstoftwo
 \else \expandafter \@secondoftwo
 \fi
}%
\providecommand \natexlab [1]{#1}%
\providecommand \enquote  [1]{``#1''}%
\providecommand \bibnamefont  [1]{#1}%
\providecommand \bibfnamefont [1]{#1}%
\providecommand \citenamefont [1]{#1}%
\providecommand \href@noop [0]{\@secondoftwo}%
\providecommand \href [0]{\begingroup \@sanitize@url \@href}%
\providecommand \@href[1]{\@@startlink{#1}\@@href}%
\providecommand \@@href[1]{\endgroup#1\@@endlink}%
\providecommand \@sanitize@url [0]{\catcode `\\12\catcode `\$12\catcode `\&12\catcode `\#12\catcode `\^12\catcode `\_12\catcode `\%12\relax}%
\providecommand \@@startlink[1]{}%
\providecommand \@@endlink[0]{}%
\providecommand \url  [0]{\begingroup\@sanitize@url \@url }%
\providecommand \@url [1]{\endgroup\@href {#1}{\urlprefix }}%
\providecommand \urlprefix  [0]{URL }%
\providecommand \Eprint [0]{\href }%
\providecommand \doibase [0]{https://doi.org/}%
\providecommand \selectlanguage [0]{\@gobble}%
\providecommand \bibinfo  [0]{\@secondoftwo}%
\providecommand \bibfield  [0]{\@secondoftwo}%
\providecommand \translation [1]{[#1]}%
\providecommand \BibitemOpen [0]{}%
\providecommand \bibitemStop [0]{}%
\providecommand \bibitemNoStop [0]{.\EOS\space}%
\providecommand \EOS [0]{\spacefactor3000\relax}%
\providecommand \BibitemShut  [1]{\csname bibitem#1\endcsname}%
\let\auto@bib@innerbib\@empty
%</preamble>
\bibitem [{\citenamefont {Daane}\ \emph {et~al.}(1954)\citenamefont {Daane}, \citenamefont {Rundle}, \citenamefont {Smith},\ and\ \citenamefont {Spedding}}]{Daane:a01231}%
  \BibitemOpen
  \bibfield  {author} {\bibinfo {author} {\bibfnamefont {A.~H.}\ \bibnamefont {Daane}}, \bibinfo {author} {\bibfnamefont {R.~E.}\ \bibnamefont {Rundle}}, \bibinfo {author} {\bibfnamefont {H.~G.}\ \bibnamefont {Smith}},\ and\ \bibinfo {author} {\bibfnamefont {F.~H.}\ \bibnamefont {Spedding}},\ }\href {https://doi.org/10.1107/S0365110X54001818} {\bibfield  {journal} {\bibinfo  {journal} {Acta Crystallographica}\ }\textbf {\bibinfo {volume} {7}},\ \bibinfo {pages} {532} (\bibinfo {year} {1954})}\BibitemShut {NoStop}%
\bibitem [{\citenamefont {Mardon}\ and\ \citenamefont {Koch}(1970)}]{MARDON1970477}%
  \BibitemOpen
  \bibfield  {author} {\bibinfo {author} {\bibfnamefont {P.}~\bibnamefont {Mardon}}\ and\ \bibinfo {author} {\bibfnamefont {C.}~\bibnamefont {Koch}},\ }\href {https://doi.org/https://doi.org/10.1016/0036-9748(70)90089-X} {\bibfield  {journal} {\bibinfo  {journal} {Scripta Metallurgica}\ }\textbf {\bibinfo {volume} {4}},\ \bibinfo {pages} {477} (\bibinfo {year} {1970})}\BibitemShut {NoStop}%
\bibitem [{\citenamefont {Coles}(1981)}]{COLES1981153}%
  \BibitemOpen
  \bibfield  {author} {\bibinfo {author} {\bibfnamefont {B.}~\bibnamefont {Coles}},\ }\href {https://doi.org/https://doi.org/10.1016/0022-5088(81)90020-5} {\bibfield  {journal} {\bibinfo  {journal} {Journal of the Less Common Metals}\ }\textbf {\bibinfo {volume} {77}},\ \bibinfo {pages} {153} (\bibinfo {year} {1981})}\BibitemShut {NoStop}%
\bibitem [{\citenamefont {Finnegan}\ \emph {et~al.}(2020)\citenamefont {Finnegan}, \citenamefont {Pace}, \citenamefont {Storm}, \citenamefont {McMahon}, \citenamefont {MacLeod}, \citenamefont {Liermann},\ and\ \citenamefont {Glazyrin}}]{PhysRevB.101.174109}%
  \BibitemOpen
  \bibfield  {author} {\bibinfo {author} {\bibfnamefont {S.~E.}\ \bibnamefont {Finnegan}}, \bibinfo {author} {\bibfnamefont {E.~J.}\ \bibnamefont {Pace}}, \bibinfo {author} {\bibfnamefont {C.~V.}\ \bibnamefont {Storm}}, \bibinfo {author} {\bibfnamefont {M.~I.}\ \bibnamefont {McMahon}}, \bibinfo {author} {\bibfnamefont {S.~G.}\ \bibnamefont {MacLeod}}, \bibinfo {author} {\bibfnamefont {H.-P.}\ \bibnamefont {Liermann}},\ and\ \bibinfo {author} {\bibfnamefont {K.}~\bibnamefont {Glazyrin}},\ }\href {https://doi.org/10.1103/PhysRevB.101.174109} {\bibfield  {journal} {\bibinfo  {journal} {Phys. Rev. B}\ }\textbf {\bibinfo {volume} {101}},\ \bibinfo {pages} {174109} (\bibinfo {year} {2020})}\BibitemShut {NoStop}%
\bibitem [{\citenamefont {Koehler}\ and\ \citenamefont {Moon}(1972)}]{PhysRevLett.29.1468}%
  \BibitemOpen
  \bibfield  {author} {\bibinfo {author} {\bibfnamefont {W.~C.}\ \bibnamefont {Koehler}}\ and\ \bibinfo {author} {\bibfnamefont {R.~M.}\ \bibnamefont {Moon}},\ }\href {https://doi.org/10.1103/PhysRevLett.29.1468} {\bibfield  {journal} {\bibinfo  {journal} {Phys. Rev. Lett.}\ }\textbf {\bibinfo {volume} {29}},\ \bibinfo {pages} {1468} (\bibinfo {year} {1972})}\BibitemShut {NoStop}%
\bibitem [{\citenamefont {Deng}\ and\ \citenamefont {Schilling}(2019)}]{PhysRevB.99.085137}%
  \BibitemOpen
  \bibfield  {author} {\bibinfo {author} {\bibfnamefont {Y.}~\bibnamefont {Deng}}\ and\ \bibinfo {author} {\bibfnamefont {J.~S.}\ \bibnamefont {Schilling}},\ }\href {https://doi.org/10.1103/PhysRevB.99.085137} {\bibfield  {journal} {\bibinfo  {journal} {Phys. Rev. B}\ }\textbf {\bibinfo {volume} {99}},\ \bibinfo {pages} {085137} (\bibinfo {year} {2019})}\BibitemShut {NoStop}%
\bibitem [{\citenamefont {Strnat}\ \emph {et~al.}(1967)\citenamefont {Strnat}, \citenamefont {Hoffer}, \citenamefont {Olson}, \citenamefont {Ostertag},\ and\ \citenamefont {Becker}}]{10.1063.1.1709459}%
  \BibitemOpen
  \bibfield  {author} {\bibinfo {author} {\bibfnamefont {K.}~\bibnamefont {Strnat}}, \bibinfo {author} {\bibfnamefont {G.}~\bibnamefont {Hoffer}}, \bibinfo {author} {\bibfnamefont {J.}~\bibnamefont {Olson}}, \bibinfo {author} {\bibfnamefont {W.}~\bibnamefont {Ostertag}},\ and\ \bibinfo {author} {\bibfnamefont {J.~J.}\ \bibnamefont {Becker}},\ }\href {https://doi.org/10.1063/1.1709459} {\bibfield  {journal} {\bibinfo  {journal} {Journal of Applied Physics}\ }\textbf {\bibinfo {volume} {38}},\ \bibinfo {pages} {1001} (\bibinfo {year} {1967})}\BibitemShut {NoStop}%
\bibitem [{\citenamefont {Min}\ \emph {et~al.}(2017)\citenamefont {Min}, \citenamefont {Goth}, \citenamefont {Lutz}, \citenamefont {Bentmann}, \citenamefont {Kang}, \citenamefont {Cho}, \citenamefont {Werner}, \citenamefont {Chen}, \citenamefont {Assaad},\ and\ \citenamefont {Reinert}}]{Min2017}%
  \BibitemOpen
  \bibfield  {author} {\bibinfo {author} {\bibfnamefont {C.-H.}\ \bibnamefont {Min}}, \bibinfo {author} {\bibfnamefont {F.}~\bibnamefont {Goth}}, \bibinfo {author} {\bibfnamefont {P.}~\bibnamefont {Lutz}}, \bibinfo {author} {\bibfnamefont {H.}~\bibnamefont {Bentmann}}, \bibinfo {author} {\bibfnamefont {B.~Y.}\ \bibnamefont {Kang}}, \bibinfo {author} {\bibfnamefont {B.~K.}\ \bibnamefont {Cho}}, \bibinfo {author} {\bibfnamefont {J.}~\bibnamefont {Werner}}, \bibinfo {author} {\bibfnamefont {K.-S.}\ \bibnamefont {Chen}}, \bibinfo {author} {\bibfnamefont {F.}~\bibnamefont {Assaad}},\ and\ \bibinfo {author} {\bibfnamefont {F.}~\bibnamefont {Reinert}},\ }\href {https://doi.org/10.1038/s41598-017-12080-5} {\bibfield  {journal} {\bibinfo  {journal} {Scientific Reports}\ }\textbf {\bibinfo {volume} {7}},\ \bibinfo {pages} {11980} (\bibinfo {year} {2017})}\BibitemShut {NoStop}%
\bibitem [{\citenamefont {Hatnean}\ \emph {et~al.}(2013)\citenamefont {Hatnean}, \citenamefont {Lees}, \citenamefont {Paul},\ and\ \citenamefont {Balakrishnan}}]{Hatnean2013}%
  \BibitemOpen
  \bibfield  {author} {\bibinfo {author} {\bibfnamefont {M.~C.}\ \bibnamefont {Hatnean}}, \bibinfo {author} {\bibfnamefont {M.~R.}\ \bibnamefont {Lees}}, \bibinfo {author} {\bibfnamefont {D.~M.}\ \bibnamefont {Paul}},\ and\ \bibinfo {author} {\bibfnamefont {G.}~\bibnamefont {Balakrishnan}},\ }\href {https://doi.org/10.1038/srep03071} {\bibfield  {journal} {\bibinfo  {journal} {Scientific Reports}\ }\textbf {\bibinfo {volume} {3}},\ \bibinfo {pages} {3071} (\bibinfo {year} {2013})}\BibitemShut {NoStop}%
\bibitem [{\citenamefont {Ruan}\ \emph {et~al.}(2014)\citenamefont {Ruan}, \citenamefont {Ye}, \citenamefont {Guo}, \citenamefont {Chen}, \citenamefont {Chen}, \citenamefont {Zhang},\ and\ \citenamefont {Wang}}]{PhysRevLett.112.136401}%
  \BibitemOpen
  \bibfield  {author} {\bibinfo {author} {\bibfnamefont {W.}~\bibnamefont {Ruan}}, \bibinfo {author} {\bibfnamefont {C.}~\bibnamefont {Ye}}, \bibinfo {author} {\bibfnamefont {M.}~\bibnamefont {Guo}}, \bibinfo {author} {\bibfnamefont {F.}~\bibnamefont {Chen}}, \bibinfo {author} {\bibfnamefont {X.}~\bibnamefont {Chen}}, \bibinfo {author} {\bibfnamefont {G.-M.}\ \bibnamefont {Zhang}},\ and\ \bibinfo {author} {\bibfnamefont {Y.}~\bibnamefont {Wang}},\ }\href {https://doi.org/10.1103/PhysRevLett.112.136401} {\bibfield  {journal} {\bibinfo  {journal} {Phys. Rev. Lett.}\ }\textbf {\bibinfo {volume} {112}},\ \bibinfo {pages} {136401} (\bibinfo {year} {2014})}\BibitemShut {NoStop}%
\bibitem [{\citenamefont {Kang}\ \emph {et~al.}(2015)\citenamefont {Kang}, \citenamefont {Choi}, \citenamefont {Kim},\ and\ \citenamefont {Min}}]{PhysRevLett.114.166404}%
  \BibitemOpen
  \bibfield  {author} {\bibinfo {author} {\bibfnamefont {C.-J.}\ \bibnamefont {Kang}}, \bibinfo {author} {\bibfnamefont {H.~C.}\ \bibnamefont {Choi}}, \bibinfo {author} {\bibfnamefont {K.}~\bibnamefont {Kim}},\ and\ \bibinfo {author} {\bibfnamefont {B.~I.}\ \bibnamefont {Min}},\ }\href {https://doi.org/10.1103/PhysRevLett.114.166404} {\bibfield  {journal} {\bibinfo  {journal} {Phys. Rev. Lett.}\ }\textbf {\bibinfo {volume} {114}},\ \bibinfo {pages} {166404} (\bibinfo {year} {2015})}\BibitemShut {NoStop}%
\bibitem [{\citenamefont {Banerjee}\ \emph {et~al.}(2022)\citenamefont {Banerjee}, \citenamefont {Plekhanov}, \citenamefont {Rungger},\ and\ \citenamefont {Weber}}]{PhysRevB.105.195135}%
  \BibitemOpen
  \bibfield  {author} {\bibinfo {author} {\bibfnamefont {D.}~\bibnamefont {Banerjee}}, \bibinfo {author} {\bibfnamefont {E.}~\bibnamefont {Plekhanov}}, \bibinfo {author} {\bibfnamefont {I.}~\bibnamefont {Rungger}},\ and\ \bibinfo {author} {\bibfnamefont {C.}~\bibnamefont {Weber}},\ }\href {https://doi.org/10.1103/PhysRevB.105.195135} {\bibfield  {journal} {\bibinfo  {journal} {Phys. Rev. B}\ }\textbf {\bibinfo {volume} {105}},\ \bibinfo {pages} {195135} (\bibinfo {year} {2022})}\BibitemShut {NoStop}%
\bibitem [{\citenamefont {Kundu}\ \emph {et~al.}(2022)\citenamefont {Kundu}, \citenamefont {Pakhira}, \citenamefont {Roy}, \citenamefont {Yilmaz}, \citenamefont {Tsujikawa}, \citenamefont {Shirai}, \citenamefont {Vescovo}, \citenamefont {Johnston}, \citenamefont {Pasupathy},\ and\ \citenamefont {Valla}}]{PhysRevB.106.245131}%
  \BibitemOpen
  \bibfield  {author} {\bibinfo {author} {\bibfnamefont {A.~K.}\ \bibnamefont {Kundu}}, \bibinfo {author} {\bibfnamefont {S.}~\bibnamefont {Pakhira}}, \bibinfo {author} {\bibfnamefont {T.}~\bibnamefont {Roy}}, \bibinfo {author} {\bibfnamefont {T.}~\bibnamefont {Yilmaz}}, \bibinfo {author} {\bibfnamefont {M.}~\bibnamefont {Tsujikawa}}, \bibinfo {author} {\bibfnamefont {M.}~\bibnamefont {Shirai}}, \bibinfo {author} {\bibfnamefont {E.}~\bibnamefont {Vescovo}}, \bibinfo {author} {\bibfnamefont {D.~C.}\ \bibnamefont {Johnston}}, \bibinfo {author} {\bibfnamefont {A.~N.}\ \bibnamefont {Pasupathy}},\ and\ \bibinfo {author} {\bibfnamefont {T.}~\bibnamefont {Valla}},\ }\href {https://doi.org/10.1103/PhysRevB.106.245131} {\bibfield  {journal} {\bibinfo  {journal} {Phys. Rev. B}\ }\textbf {\bibinfo {volume} {106}},\ \bibinfo {pages} {245131} (\bibinfo {year} {2022})}\BibitemShut {NoStop}%
\bibitem [{\citenamefont {Z\"olfl}\ \emph {et~al.}(2001)\citenamefont {Z\"olfl}, \citenamefont {Nekrasov}, \citenamefont {Pruschke}, \citenamefont {Anisimov},\ and\ \citenamefont {Keller}}]{PhysRevLett.87.276403}%
  \BibitemOpen
  \bibfield  {author} {\bibinfo {author} {\bibfnamefont {M.~B.}\ \bibnamefont {Z\"olfl}}, \bibinfo {author} {\bibfnamefont {I.~A.}\ \bibnamefont {Nekrasov}}, \bibinfo {author} {\bibfnamefont {T.}~\bibnamefont {Pruschke}}, \bibinfo {author} {\bibfnamefont {V.~I.}\ \bibnamefont {Anisimov}},\ and\ \bibinfo {author} {\bibfnamefont {J.}~\bibnamefont {Keller}},\ }\href {https://doi.org/10.1103/PhysRevLett.87.276403} {\bibfield  {journal} {\bibinfo  {journal} {Phys. Rev. Lett.}\ }\textbf {\bibinfo {volume} {87}},\ \bibinfo {pages} {276403} (\bibinfo {year} {2001})}\BibitemShut {NoStop}%
\bibitem [{\citenamefont {Held}\ \emph {et~al.}(2001)\citenamefont {Held}, \citenamefont {McMahan},\ and\ \citenamefont {Scalettar}}]{PhysRevLett.87.276404}%
  \BibitemOpen
  \bibfield  {author} {\bibinfo {author} {\bibfnamefont {K.}~\bibnamefont {Held}}, \bibinfo {author} {\bibfnamefont {A.~K.}\ \bibnamefont {McMahan}},\ and\ \bibinfo {author} {\bibfnamefont {R.~T.}\ \bibnamefont {Scalettar}},\ }\href {https://doi.org/10.1103/PhysRevLett.87.276404} {\bibfield  {journal} {\bibinfo  {journal} {Phys. Rev. Lett.}\ }\textbf {\bibinfo {volume} {87}},\ \bibinfo {pages} {276404} (\bibinfo {year} {2001})}\BibitemShut {NoStop}%
\bibitem [{\citenamefont {Huang}\ and\ \citenamefont {Lu}(2019)}]{PhysRevB.99.045122}%
  \BibitemOpen
  \bibfield  {author} {\bibinfo {author} {\bibfnamefont {L.}~\bibnamefont {Huang}}\ and\ \bibinfo {author} {\bibfnamefont {H.}~\bibnamefont {Lu}},\ }\href {https://doi.org/10.1103/PhysRevB.99.045122} {\bibfield  {journal} {\bibinfo  {journal} {Phys. Rev. B}\ }\textbf {\bibinfo {volume} {99}},\ \bibinfo {pages} {045122} (\bibinfo {year} {2019})}\BibitemShut {NoStop}%
\bibitem [{\citenamefont {McMahan}\ \emph {et~al.}(2009)\citenamefont {McMahan}, \citenamefont {Scalettar},\ and\ \citenamefont {Jarrell}}]{PhysRevB.80.235105}%
  \BibitemOpen
  \bibfield  {author} {\bibinfo {author} {\bibfnamefont {A.~K.}\ \bibnamefont {McMahan}}, \bibinfo {author} {\bibfnamefont {R.~T.}\ \bibnamefont {Scalettar}},\ and\ \bibinfo {author} {\bibfnamefont {M.}~\bibnamefont {Jarrell}},\ }\href {https://doi.org/10.1103/PhysRevB.80.235105} {\bibfield  {journal} {\bibinfo  {journal} {Phys. Rev. B}\ }\textbf {\bibinfo {volume} {80}},\ \bibinfo {pages} {235105} (\bibinfo {year} {2009})}\BibitemShut {NoStop}%
\bibitem [{\citenamefont {Shim}\ \emph {et~al.}(2007)\citenamefont {Shim}, \citenamefont {Haule},\ and\ \citenamefont {Kotliar}}]{Shim2007}%
  \BibitemOpen
  \bibfield  {author} {\bibinfo {author} {\bibfnamefont {J.~H.}\ \bibnamefont {Shim}}, \bibinfo {author} {\bibfnamefont {K.}~\bibnamefont {Haule}},\ and\ \bibinfo {author} {\bibfnamefont {G.}~\bibnamefont {Kotliar}},\ }\href {https://doi.org/10.1038/nature05647} {\bibfield  {journal} {\bibinfo  {journal} {Nature}\ }\textbf {\bibinfo {volume} {446}},\ \bibinfo {pages} {513} (\bibinfo {year} {2007})}\BibitemShut {NoStop}%
\bibitem [{\citenamefont {Amadon}(2016)}]{PhysRevB.94.115148}%
  \BibitemOpen
  \bibfield  {author} {\bibinfo {author} {\bibfnamefont {B.}~\bibnamefont {Amadon}},\ }\href {https://doi.org/10.1103/PhysRevB.94.115148} {\bibfield  {journal} {\bibinfo  {journal} {Phys. Rev. B}\ }\textbf {\bibinfo {volume} {94}},\ \bibinfo {pages} {115148} (\bibinfo {year} {2016})}\BibitemShut {NoStop}%
\bibitem [{\citenamefont {Huang}\ and\ \citenamefont {Lu}(2020)}]{PhysRevB.101.125123}%
  \BibitemOpen
  \bibfield  {author} {\bibinfo {author} {\bibfnamefont {L.}~\bibnamefont {Huang}}\ and\ \bibinfo {author} {\bibfnamefont {H.}~\bibnamefont {Lu}},\ }\href {https://doi.org/10.1103/PhysRevB.101.125123} {\bibfield  {journal} {\bibinfo  {journal} {Phys. Rev. B}\ }\textbf {\bibinfo {volume} {101}},\ \bibinfo {pages} {125123} (\bibinfo {year} {2020})}\BibitemShut {NoStop}%
\bibitem [{\citenamefont {Kotliar}\ \emph {et~al.}(2006)\citenamefont {Kotliar}, \citenamefont {Savrasov}, \citenamefont {Haule}, \citenamefont {Oudovenko}, \citenamefont {Parcollet},\ and\ \citenamefont {Marianetti}}]{RevModPhys.78.865}%
  \BibitemOpen
  \bibfield  {author} {\bibinfo {author} {\bibfnamefont {G.}~\bibnamefont {Kotliar}}, \bibinfo {author} {\bibfnamefont {S.~Y.}\ \bibnamefont {Savrasov}}, \bibinfo {author} {\bibfnamefont {K.}~\bibnamefont {Haule}}, \bibinfo {author} {\bibfnamefont {V.~S.}\ \bibnamefont {Oudovenko}}, \bibinfo {author} {\bibfnamefont {O.}~\bibnamefont {Parcollet}},\ and\ \bibinfo {author} {\bibfnamefont {C.~A.}\ \bibnamefont {Marianetti}},\ }\href {https://doi.org/10.1103/RevModPhys.78.865} {\bibfield  {journal} {\bibinfo  {journal} {Rev. Mod. Phys.}\ }\textbf {\bibinfo {volume} {78}},\ \bibinfo {pages} {865} (\bibinfo {year} {2006})}\BibitemShut {NoStop}%
\bibitem [{\citenamefont {Anisimov}\ \emph {et~al.}(1997{\natexlab{a}})\citenamefont {Anisimov}, \citenamefont {Poteryaev}, \citenamefont {Korotin}, \citenamefont {Anokhin},\ and\ \citenamefont {Kotliar}}]{VAnisimov1997}%
  \BibitemOpen
  \bibfield  {author} {\bibinfo {author} {\bibfnamefont {V.~I.}\ \bibnamefont {Anisimov}}, \bibinfo {author} {\bibfnamefont {A.~I.}\ \bibnamefont {Poteryaev}}, \bibinfo {author} {\bibfnamefont {M.~A.}\ \bibnamefont {Korotin}}, \bibinfo {author} {\bibfnamefont {A.~O.}\ \bibnamefont {Anokhin}},\ and\ \bibinfo {author} {\bibfnamefont {G.}~\bibnamefont {Kotliar}},\ }\href {https://doi.org/10.1088/0953-8984/9/35/010} {\bibfield  {journal} {\bibinfo  {journal} {Journal of Physics: Condensed Matter}\ }\textbf {\bibinfo {volume} {9}},\ \bibinfo {pages} {7359} (\bibinfo {year} {1997}{\natexlab{a}})}\BibitemShut {NoStop}%
\bibitem [{\citenamefont {Lichtenstein}\ and\ \citenamefont {Katsnelson}(1998)}]{PhysRevB.57.6884}%
  \BibitemOpen
  \bibfield  {author} {\bibinfo {author} {\bibfnamefont {A.~I.}\ \bibnamefont {Lichtenstein}}\ and\ \bibinfo {author} {\bibfnamefont {M.~I.}\ \bibnamefont {Katsnelson}},\ }\href {https://doi.org/10.1103/PhysRevB.57.6884} {\bibfield  {journal} {\bibinfo  {journal} {Phys. Rev. B}\ }\textbf {\bibinfo {volume} {57}},\ \bibinfo {pages} {6884} (\bibinfo {year} {1998})}\BibitemShut {NoStop}%
\bibitem [{\citenamefont {Georges}\ \emph {et~al.}(1996)\citenamefont {Georges}, \citenamefont {Kotliar}, \citenamefont {Krauth},\ and\ \citenamefont {Rozenberg}}]{RevModPhys.68.13}%
  \BibitemOpen
  \bibfield  {author} {\bibinfo {author} {\bibfnamefont {A.}~\bibnamefont {Georges}}, \bibinfo {author} {\bibfnamefont {G.}~\bibnamefont {Kotliar}}, \bibinfo {author} {\bibfnamefont {W.}~\bibnamefont {Krauth}},\ and\ \bibinfo {author} {\bibfnamefont {M.~J.}\ \bibnamefont {Rozenberg}},\ }\href {https://doi.org/10.1103/RevModPhys.68.13} {\bibfield  {journal} {\bibinfo  {journal} {Rev. Mod. Phys.}\ }\textbf {\bibinfo {volume} {68}},\ \bibinfo {pages} {13} (\bibinfo {year} {1996})}\BibitemShut {NoStop}%
\bibitem [{\citenamefont {Metzner}\ and\ \citenamefont {Vollhardt}(1989)}]{PhysRevLett.62.324}%
  \BibitemOpen
  \bibfield  {author} {\bibinfo {author} {\bibfnamefont {W.}~\bibnamefont {Metzner}}\ and\ \bibinfo {author} {\bibfnamefont {D.}~\bibnamefont {Vollhardt}},\ }\href {https://doi.org/10.1103/PhysRevLett.62.324} {\bibfield  {journal} {\bibinfo  {journal} {Phys. Rev. Lett.}\ }\textbf {\bibinfo {volume} {62}},\ \bibinfo {pages} {324} (\bibinfo {year} {1989})}\BibitemShut {NoStop}%
\bibitem [{\citenamefont {M{\"u}ller-Hartmann}(1989)}]{MllerHartmann1989}%
  \BibitemOpen
  \bibfield  {author} {\bibinfo {author} {\bibfnamefont {E.}~\bibnamefont {M{\"u}ller-Hartmann}},\ }\href {https://doi.org/10.1007/BF01311397} {\bibfield  {journal} {\bibinfo  {journal} {Zeitschrift f{\"u}r Physik B Condensed Matter}\ }\textbf {\bibinfo {volume} {74}},\ \bibinfo {pages} {507} (\bibinfo {year} {1989})}\BibitemShut {NoStop}%
\bibitem [{\citenamefont {Blaha}\ \emph {et~al.}(2001)\citenamefont {Blaha}, \citenamefont {Schwarz}, \citenamefont {Madsen}, \citenamefont {Kvasnicka},\ and\ \citenamefont {Luitz}}]{wien2k}%
  \BibitemOpen
  \bibfield  {author} {\bibinfo {author} {\bibfnamefont {P.}~\bibnamefont {Blaha}}, \bibinfo {author} {\bibfnamefont {K.}~\bibnamefont {Schwarz}}, \bibinfo {author} {\bibfnamefont {G.}~\bibnamefont {Madsen}}, \bibinfo {author} {\bibfnamefont {D.}~\bibnamefont {Kvasnicka}},\ and\ \bibinfo {author} {\bibfnamefont {J.}~\bibnamefont {Luitz}},\ }\href@noop {} {\bibfield  {journal} {\bibinfo  {journal} {Technische Universität Wien, Wien}\ }\textbf {\bibinfo {volume} {28}} (\bibinfo {year} {2001})}\BibitemShut {NoStop}%
\bibitem [{\citenamefont {Haule}\ \emph {et~al.}(2010)\citenamefont {Haule}, \citenamefont {Yee},\ and\ \citenamefont {Kim}}]{PhysRevB.81.195107}%
  \BibitemOpen
  \bibfield  {author} {\bibinfo {author} {\bibfnamefont {K.}~\bibnamefont {Haule}}, \bibinfo {author} {\bibfnamefont {C.-H.}\ \bibnamefont {Yee}},\ and\ \bibinfo {author} {\bibfnamefont {K.}~\bibnamefont {Kim}},\ }\href {https://doi.org/10.1103/PhysRevB.81.195107} {\bibfield  {journal} {\bibinfo  {journal} {Phys. Rev. B}\ }\textbf {\bibinfo {volume} {81}},\ \bibinfo {pages} {195107} (\bibinfo {year} {2010})}\BibitemShut {NoStop}%
\bibitem [{\citenamefont {Hinuma}\ \emph {et~al.}(2017)\citenamefont {Hinuma}, \citenamefont {Pizzi}, \citenamefont {Kumagai}, \citenamefont {Oba},\ and\ \citenamefont {Tanaka}}]{HINUMA2017140}%
  \BibitemOpen
  \bibfield  {author} {\bibinfo {author} {\bibfnamefont {Y.}~\bibnamefont {Hinuma}}, \bibinfo {author} {\bibfnamefont {G.}~\bibnamefont {Pizzi}}, \bibinfo {author} {\bibfnamefont {Y.}~\bibnamefont {Kumagai}}, \bibinfo {author} {\bibfnamefont {F.}~\bibnamefont {Oba}},\ and\ \bibinfo {author} {\bibfnamefont {I.}~\bibnamefont {Tanaka}},\ }\href {https://doi.org/https://doi.org/10.1016/j.commatsci.2016.10.015} {\bibfield  {journal} {\bibinfo  {journal} {Computational Materials Science}\ }\textbf {\bibinfo {volume} {128}},\ \bibinfo {pages} {140} (\bibinfo {year} {2017})}\BibitemShut {NoStop}%
\bibitem [{\citenamefont {Togo}\ \emph {et~al.}(2024)\citenamefont {Togo}, \citenamefont {Shinohara},\ and\ \citenamefont {Tanaka}}]{togo2024}%
  \BibitemOpen
  \bibfield  {author} {\bibinfo {author} {\bibfnamefont {A.}~\bibnamefont {Togo}}, \bibinfo {author} {\bibfnamefont {K.}~\bibnamefont {Shinohara}},\ and\ \bibinfo {author} {\bibfnamefont {I.}~\bibnamefont {Tanaka}},\ }\href {https://arxiv.org/abs/1808.01590} {\bibinfo {title} {$\texttt{Spglib}$: a software library for crystal symmetry search}} (\bibinfo {year} {2024}),\ \Eprint {https://arxiv.org/abs/1808.01590} {arXiv:1808.01590 [cond-mat.mtrl-sci]} \BibitemShut {NoStop}%
\bibitem [{\citenamefont {Kumar}\ and\ \citenamefont {Srivastava}(1969)}]{Kumar:a07033}%
  \BibitemOpen
  \bibfield  {author} {\bibinfo {author} {\bibfnamefont {J.}~\bibnamefont {Kumar}}\ and\ \bibinfo {author} {\bibfnamefont {O.~N.}\ \bibnamefont {Srivastava}},\ }\href {https://doi.org/10.1107/S0567740869006212} {\bibfield  {journal} {\bibinfo  {journal} {Acta Crystallographica Section B}\ }\textbf {\bibinfo {volume} {25}},\ \bibinfo {pages} {2654} (\bibinfo {year} {1969})}\BibitemShut {NoStop}%
\bibitem [{\citenamefont {Kresse}\ and\ \citenamefont {Furthm\"uller}(1996)}]{PhysRevB.54.11169}%
  \BibitemOpen
  \bibfield  {author} {\bibinfo {author} {\bibfnamefont {G.}~\bibnamefont {Kresse}}\ and\ \bibinfo {author} {\bibfnamefont {J.}~\bibnamefont {Furthm\"uller}},\ }\href {https://doi.org/10.1103/PhysRevB.54.11169} {\bibfield  {journal} {\bibinfo  {journal} {Phys. Rev. B}\ }\textbf {\bibinfo {volume} {54}},\ \bibinfo {pages} {11169} (\bibinfo {year} {1996})}\BibitemShut {NoStop}%
\bibitem [{\citenamefont {Perdew}\ \emph {et~al.}(1996)\citenamefont {Perdew}, \citenamefont {Burke},\ and\ \citenamefont {Ernzerhof}}]{PhysRevLett.77.3865}%
  \BibitemOpen
  \bibfield  {author} {\bibinfo {author} {\bibfnamefont {J.~P.}\ \bibnamefont {Perdew}}, \bibinfo {author} {\bibfnamefont {K.}~\bibnamefont {Burke}},\ and\ \bibinfo {author} {\bibfnamefont {M.}~\bibnamefont {Ernzerhof}},\ }\href {https://doi.org/10.1103/PhysRevLett.77.3865} {\bibfield  {journal} {\bibinfo  {journal} {Phys. Rev. Lett.}\ }\textbf {\bibinfo {volume} {77}},\ \bibinfo {pages} {3865} (\bibinfo {year} {1996})}\BibitemShut {NoStop}%
\bibitem [{\citenamefont {Werner}\ \emph {et~al.}(2006)\citenamefont {Werner}, \citenamefont {Comanac}, \citenamefont {de' Medici}, \citenamefont {Troyer},\ and\ \citenamefont {Millis}}]{PhysRevLett.97.076405}%
  \BibitemOpen
  \bibfield  {author} {\bibinfo {author} {\bibfnamefont {P.}~\bibnamefont {Werner}}, \bibinfo {author} {\bibfnamefont {A.}~\bibnamefont {Comanac}}, \bibinfo {author} {\bibfnamefont {L.}~\bibnamefont {de' Medici}}, \bibinfo {author} {\bibfnamefont {M.}~\bibnamefont {Troyer}},\ and\ \bibinfo {author} {\bibfnamefont {A.~J.}\ \bibnamefont {Millis}},\ }\href {https://doi.org/10.1103/PhysRevLett.97.076405} {\bibfield  {journal} {\bibinfo  {journal} {Phys. Rev. Lett.}\ }\textbf {\bibinfo {volume} {97}},\ \bibinfo {pages} {076405} (\bibinfo {year} {2006})}\BibitemShut {NoStop}%
\bibitem [{\citenamefont {Haule}(2007)}]{PhysRevB.75.155113}%
  \BibitemOpen
  \bibfield  {author} {\bibinfo {author} {\bibfnamefont {K.}~\bibnamefont {Haule}},\ }\href {https://doi.org/10.1103/PhysRevB.75.155113} {\bibfield  {journal} {\bibinfo  {journal} {Phys. Rev. B}\ }\textbf {\bibinfo {volume} {75}},\ \bibinfo {pages} {155113} (\bibinfo {year} {2007})}\BibitemShut {NoStop}%
\bibitem [{\citenamefont {Gull}\ \emph {et~al.}(2011)\citenamefont {Gull}, \citenamefont {Millis}, \citenamefont {Lichtenstein}, \citenamefont {Rubtsov}, \citenamefont {Troyer},\ and\ \citenamefont {Werner}}]{RevModPhys.83.349}%
  \BibitemOpen
  \bibfield  {author} {\bibinfo {author} {\bibfnamefont {E.}~\bibnamefont {Gull}}, \bibinfo {author} {\bibfnamefont {A.~J.}\ \bibnamefont {Millis}}, \bibinfo {author} {\bibfnamefont {A.~I.}\ \bibnamefont {Lichtenstein}}, \bibinfo {author} {\bibfnamefont {A.~N.}\ \bibnamefont {Rubtsov}}, \bibinfo {author} {\bibfnamefont {M.}~\bibnamefont {Troyer}},\ and\ \bibinfo {author} {\bibfnamefont {P.}~\bibnamefont {Werner}},\ }\href {https://doi.org/10.1103/RevModPhys.83.349} {\bibfield  {journal} {\bibinfo  {journal} {Rev. Mod. Phys.}\ }\textbf {\bibinfo {volume} {83}},\ \bibinfo {pages} {349} (\bibinfo {year} {2011})}\BibitemShut {NoStop}%
\bibitem [{\citenamefont {Plekhanov}\ \emph {et~al.}(2018)\citenamefont {Plekhanov}, \citenamefont {Hasnip}, \citenamefont {Sacksteder}, \citenamefont {Probert}, \citenamefont {Clark}, \citenamefont {Refson},\ and\ \citenamefont {Weber}}]{PhysRevB.98.075129}%
  \BibitemOpen
  \bibfield  {author} {\bibinfo {author} {\bibfnamefont {E.}~\bibnamefont {Plekhanov}}, \bibinfo {author} {\bibfnamefont {P.}~\bibnamefont {Hasnip}}, \bibinfo {author} {\bibfnamefont {V.}~\bibnamefont {Sacksteder}}, \bibinfo {author} {\bibfnamefont {M.}~\bibnamefont {Probert}}, \bibinfo {author} {\bibfnamefont {S.~J.}\ \bibnamefont {Clark}}, \bibinfo {author} {\bibfnamefont {K.}~\bibnamefont {Refson}},\ and\ \bibinfo {author} {\bibfnamefont {C.}~\bibnamefont {Weber}},\ }\href {https://doi.org/10.1103/PhysRevB.98.075129} {\bibfield  {journal} {\bibinfo  {journal} {Phys. Rev. B}\ }\textbf {\bibinfo {volume} {98}},\ \bibinfo {pages} {075129} (\bibinfo {year} {2018})}\BibitemShut {NoStop}%
\bibitem [{\citenamefont {Jarrell}\ and\ \citenamefont {Gubernatis}(1996)}]{JARRELL1996133}%
  \BibitemOpen
  \bibfield  {author} {\bibinfo {author} {\bibfnamefont {M.}~\bibnamefont {Jarrell}}\ and\ \bibinfo {author} {\bibfnamefont {J.}~\bibnamefont {Gubernatis}},\ }\href {https://doi.org/https://doi.org/10.1016/0370-1573(95)00074-7} {\bibfield  {journal} {\bibinfo  {journal} {Physics Reports}\ }\textbf {\bibinfo {volume} {269}},\ \bibinfo {pages} {133} (\bibinfo {year} {1996})}\BibitemShut {NoStop}%
\bibitem [{\citenamefont {Doniach}(1977)}]{DONIACH1977231}%
  \BibitemOpen
  \bibfield  {author} {\bibinfo {author} {\bibfnamefont {S.}~\bibnamefont {Doniach}},\ }\href {https://doi.org/https://doi.org/10.1016/0378-4363(77)90190-5} {\bibfield  {journal} {\bibinfo  {journal} {Physica B+C}\ }\textbf {\bibinfo {volume} {91}},\ \bibinfo {pages} {231} (\bibinfo {year} {1977})}\BibitemShut {NoStop}%
\bibitem [{\citenamefont {Brod{\'e}n}(1972)}]{Brodén1972}%
  \BibitemOpen
  \bibfield  {author} {\bibinfo {author} {\bibfnamefont {G.}~\bibnamefont {Brod{\'e}n}},\ }\href {https://doi.org/10.1007/BF02422677} {\bibfield  {journal} {\bibinfo  {journal} {Physik der kondensierten Materie}\ }\textbf {\bibinfo {volume} {15}},\ \bibinfo {pages} {171} (\bibinfo {year} {1972})}\BibitemShut {NoStop}%
\bibitem [{\citenamefont {Lang}\ and\ \citenamefont {Baer}(1979)}]{LANG1979945}%
  \BibitemOpen
  \bibfield  {author} {\bibinfo {author} {\bibfnamefont {J.}~\bibnamefont {Lang}}\ and\ \bibinfo {author} {\bibfnamefont {Y.}~\bibnamefont {Baer}},\ }\href {https://doi.org/https://doi.org/10.1016/0038-1098(79)90006-1} {\bibfield  {journal} {\bibinfo  {journal} {Solid State Communications}\ }\textbf {\bibinfo {volume} {31}},\ \bibinfo {pages} {945} (\bibinfo {year} {1979})}\BibitemShut {NoStop}%
\bibitem [{\citenamefont {Speier}\ \emph {et~al.}(1984)\citenamefont {Speier}, \citenamefont {Fuggle}, \citenamefont {Zeller}, \citenamefont {Ackermann}, \citenamefont {Szot}, \citenamefont {Hillebrecht},\ and\ \citenamefont {Campagna}}]{PhysRevB.30.6921}%
  \BibitemOpen
  \bibfield  {author} {\bibinfo {author} {\bibfnamefont {W.}~\bibnamefont {Speier}}, \bibinfo {author} {\bibfnamefont {J.~C.}\ \bibnamefont {Fuggle}}, \bibinfo {author} {\bibfnamefont {R.}~\bibnamefont {Zeller}}, \bibinfo {author} {\bibfnamefont {B.}~\bibnamefont {Ackermann}}, \bibinfo {author} {\bibfnamefont {K.}~\bibnamefont {Szot}}, \bibinfo {author} {\bibfnamefont {F.~U.}\ \bibnamefont {Hillebrecht}},\ and\ \bibinfo {author} {\bibfnamefont {M.}~\bibnamefont {Campagna}},\ }\href {https://doi.org/10.1103/PhysRevB.30.6921} {\bibfield  {journal} {\bibinfo  {journal} {Phys. Rev. B}\ }\textbf {\bibinfo {volume} {30}},\ \bibinfo {pages} {6921} (\bibinfo {year} {1984})}\BibitemShut {NoStop}%
\bibitem [{\citenamefont {Anisimov}\ \emph {et~al.}(1997{\natexlab{b}})\citenamefont {Anisimov}, \citenamefont {Aryasetiawan},\ and\ \citenamefont {Lichtenstein}}]{Anisimov_1997}%
  \BibitemOpen
  \bibfield  {author} {\bibinfo {author} {\bibfnamefont {V.~I.}\ \bibnamefont {Anisimov}}, \bibinfo {author} {\bibfnamefont {F.}~\bibnamefont {Aryasetiawan}},\ and\ \bibinfo {author} {\bibfnamefont {A.~I.}\ \bibnamefont {Lichtenstein}},\ }\href {https://doi.org/10.1088/0953-8984/9/4/002} {\bibfield  {journal} {\bibinfo  {journal} {Journal of Physics: Condensed Matter}\ }\textbf {\bibinfo {volume} {9}},\ \bibinfo {pages} {767} (\bibinfo {year} {1997}{\natexlab{b}})}\BibitemShut {NoStop}%
\end{thebibliography}%

\end{document}